\documentclass[final,4p,times]{elsarticle}
\usepackage{wrapfig}
\usepackage{graphics}
\usepackage{graphicx}
\usepackage{epsfig}
\usepackage{amssymb}
\usepackage{amsthm}
\usepackage{amsmath}

\usepackage{caption}
\usepackage[labelfont=bf]{caption}
\usepackage[labelsep=space]{caption}
\usepackage{xcolor}
\usepackage{hyperref}
\hypersetup{
  colorlinks,
  citecolor=[violet],
  linkcolor=[red],
  urlcolor=[blue]}

\journal{Physica B: Condensed Matter}

\begin{document}

\begin{frontmatter}

\title{Structure and magnetic properties of a La$_{0.75}$Sr$_{0.25}$Cr$_{0.90}$O$_{3-\delta}$ single crystal}

\author[1]{Kaitong Sun\fnref{fn1}}
\author[1]{Yinghao Zhu\fnref{fn1}}
\fntext[fn1]{These authors contributed equally.}
\author[2]{S. Yano} %()}
\author[1]{Qian Zhao} %()}
\author[1]{Muqing Su} %()}
\author[1]{Guanping Xu} %()}
\author[1]{Ruifeng Zheng} %()}
\author[1]{Ying Ellie Fu} %()}
\author[1]{Hai-Feng Li\corref{cor1}}
\cortext[cor1]{Corresponding author}
\ead{haifengli@um.edu.mo}
\affiliation[1]{organization={Institute of Applied Physics and Materials Engineering, University of Macau},
            addressline={Avenida da Universidade, Taipa},
            city={Macao SAR},
            postcode={999078},
     %       state={},
            country={China}}
\affiliation[2]{organization={National Synchrotron Radiation Research Center, Neutron Group},
            city={Hsinchu},
            postcode={30077},
            country={Taiwan}}

\begin{abstract}
We have successfully grown large and good-quality single crystals of the \\ La$_{0.75}$Sr$_{0.25}$Cr$_{0.90}$O$_{3-\delta}$ compound using the floating-zone method with laser diodes. We investigated the crystal quality, crystallography, chemical composition, magnetic properties and the oxidation state of Cr in the grown single crystals by employing a combination of techniques, including X-ray Laue and powder diffraction, scanning electron microscopy, magnetization measurements, X-ray photoelectron spectroscopy and light absorption. The La$_{0.75}$Sr$_{0.25}$Cr$_{0.90}$O$_{3-\delta}$ single crystal exhibits a single-phase composition, crystallizing in a trigonal structure with the space group $R\bar{3}c$ at room temperature. The chemical composition was determined as La$_{0.75}$Sr$_{0.25}$Cr$_{0.90}$O$_{3-\delta}$, indicating a significant chromium deficiency. Upon warming, we observed five distinctive characteristic temperatures, namely $T_1 =$ 21.50(1) K, $T_2 =$ 34.98(1) K, $T_3 =$ 117.94(1) K, $T_4 =$ 155.01(1) K, and $T_{\textrm{N}} =$ 271.80(1) K, revealing five distinct magnetic anomalies. Our magnetization study allows us to explore the nature of these anomalies. Remarkably, the oxidation state of chromium in the single-crystal La$_{0.75}$Sr$_{0.25}$Cr$_{0.90}$O$_{3-\delta}$, characterized by a band gap of 1.630(8) eV, is exclusively attributed to Cr$^{3+}$ ions, making a departure from the findings of previous studies on polycrystalline materials.
\end{abstract}

\begin{graphicalabstract}
\includegraphics[width=0.88\textwidth]{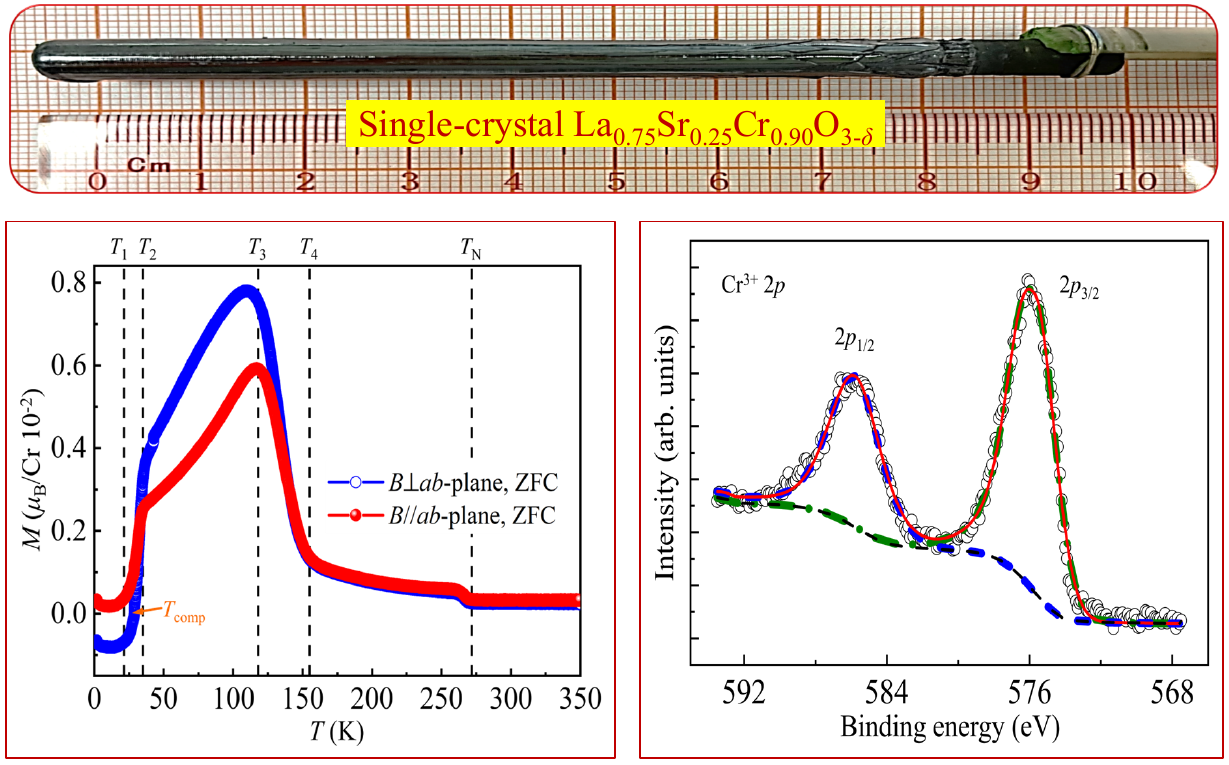}
\end{graphicalabstract}

\begin{graphicalabstract}
\includegraphics[width=0.88\textwidth]{Graphical-Abstract}\newline
\noindent{\textbf{Caption of Graphical Abstract:} We present a La$_{0.75}$Sr$_{0.25}$Cr$_{0.90}$O$_{3-\delta}$ single crystal that we successfully grew, along with magnetization data revealing five distinct features and XPS measurements confirming the presence of a solely Cr$^{3+}$ oxidation state.}
\end{graphicalabstract}

%Research highlights
\begin{highlights}
\item We have successfully grown single crystals of La$_{0.75}$Sr$_{0.25}$Cr$_{0.90}$O$_{3-\delta}$.
\item The room-temperature structure of La$_{0.75}$Sr$_{0.25}$Cr$_{0.90}$O$_{3-\delta}$ was identified as trigonal.
\item We observed five distinct magnetic anomalies in La$_{0.75}$Sr$_{0.25}$Cr$_{0.90}$O$_{3-\delta}$.
\item We have discussed the nature of these magnetic anomalies.
\item The oxidation state of chromium in La$_{0.75}$Sr$_{0.25}$Cr$_{0.90}$O$_{3-\delta}$ is solely +3.

\end{highlights}


\end{frontmatter}


\section{Introduction}
\label{}

The La$_{0.75}$Sr$_{0.25}$CrO$_3$ compound belongs to the family of perovskite-type rare-earth orthochromates. This material has garnered significant attention for its potential applications in various fields, including catalysis \cite{tan2013catalytic}, solid oxide fuel cells \cite{zheng2009cr}, sensing, thin-film electrodes \cite{liu2022improvement,machado2023chemical}, and magnetic devices \cite{ZHU2023214873}.

A neutron powder diffraction study \cite{chakraborty2006structural} conducted at high temperatures (300--1400 K) on polycrystalline La$_{0.75}$Sr$_{0.25}$CrO$_3$ samples synthesized through the solid-state route has unveiled the impact of Sr doping. It reduces the structural transition temperature, shifting it from orthorhombic to rhombohedral. Consequently, La$_{0.75}$Sr$_{0.25}$CrO$_3$ exhibits a multiphase character, with room-temperature coexistence of two phases: 89(1) wt$\%$ orthorhombic and a fraction of 11(1) wt$\%$ rhombohedral. At elevated temperatures (480--1400 K), a three-phase coexistence, including cubic, was observed. Another study \cite{yusuf2008magnetic} on polycrystalline La$_{0.75}$Sr$_{0.25}$CrO$_3$ samples demonstrates the presence of three phases at room temperature: orthorhombic, rhombohedral, and cubic. In contrast, oxygen-deficient La$_{0.75}$Sr$_{0.25}$CrO$_3$ polycrystals crystallize into rhombohedral symmetry only at room temperature \cite{khattak1977structural}. Furthermore, La$_{1-x}$Sr$_{x}$CrO$_3$ ($x =$ 0.0--0.5) compounds prepared using the sol-gel method at room temperature exhibit orthorhombic perovskite GdFeO$_3$-type structures \cite{liu2000mixed}. These polycrystalline samples show different structural phases at room temperature, likely influenced by varying synthesis conditions. Theoretically, the structure of substituted perovskite compounds can be adjusted by tuning the tolerance factor \cite{li2008synthesis}. In the La$_{0.75}$Sr$_{0.25}$CrO$_3$ compound, Sr$^{2+}$ ions are introduced into La$^{3+}$ sites at regular intervals, reducing the valence state of cations and thus resulting in the presence of Cr$^{4+}$ ions to maintain electrical neutrality \cite{karim1979localized}.

Neutron powder diffraction measurements were conducted on the magnetic structure of the polycrystalline La$_{0.75}$Sr$_{0.25}$CrO$_3$ compound. Above 250 K, the magnetic Bragg peak (101) (in the $Pbnm$ symmetry) disappears upon heating, leading to the determination that the magnetic structure of La$_{0.75}$Sr$_{0.25}$CrO$_3$ assumes a $G$-type antiferromagnetic (AFM) structure. This structure exhibits spin alignments along the crystallographic \emph{b} axis (in the $Pbnm$ symmetry) with 75 at\% of Cr$^{3+}$ (2.25 $\mu_\textrm{B}$) and 25 at\% of Cr$^{4+}$ (0.50 $\mu_\textrm{B}$) at low temperatures. Notably, there is no spin reorientation transition observed below 300 K \cite{yusuf2008magnetic,chakraborty2008magnetic}.

A core-level analysis of Cr ions, conducted through X-ray photoelectron spectroscopy measurements on La$_{1-x}$Sr$_{x}$CrO$_3$ ($x =$ 0.0--0.5) compounds, reveals that Sr doping induces a transformation in the valence state of Cr ions, transitioning from Cr$^{3+}$ to Cr$^{6+}$ with increasing levels of doping. It is noteworthy that Cr$^{4+}$ ions, characterized by a binding energy of 577.5 eV, were not observed in this study \cite{liu2000mixed}.

Most of the previous studies on La$_{0.75}$Sr$_{0.25}$CrO$_3$ have primarily focused on polycrystalline samples \cite{chakraborty2006structural,yusuf2008magnetic,khattak1977structural,liu2000mixed,karim1979localized,chakraborty2008magnetic,devi1992preparation}. To address the existing controversies and offer more reliable insights into the crystalline and magnetic properties of the La$_{0.75}$Sr$_{0.25}$CrO$_3$ compound, it is imperative to work with large and high-quality single crystals. In this study, we have achieved the successful growth of La$_{0.75}$Sr$_{0.25}$Cr$_{0.90}$O$_{3-\delta}$ single crystals using the laser-diode floating-zone method. Our comprehensive in-house characterization has encompassed the structural phase, crystal quality, chemical composition, magnetic properties, electronic band gap and the oxidation state of Cr in a La$_{0.75}$Sr$_{0.25}$Cr$_{0.90}$O$_{3-\delta}$ single crystal.

\section{Experimental}

Polycrystalline La$_{0.75}$Sr$_{0.25}$CrO$_3$ samples were prepared using the traditional solid-state reaction method \cite{li2008synthesis,li2009crystal,li2007neutron}. Stoichiometric amounts of raw materials, including La$_2$O$_3$ (Alfa Aesar, 99.9$\%$), SrCO$_3$ (Alfa Aesar, 99.99$\%$), and Cr$_2$O$_3$ (Alfa Aesar, 99.6$\%$), were thoroughly mixed and then mechanically milled using a Vibratory Micro Mill (FRITSCH PULVERISETTE 0). The resulting mixture was subjected to calcination at 1000 $^{\circ}$C for 24 h and was subsequently reground for 1 h. A second calcination step at 1100 $^{\circ}$C for 24 h ensured the formation of a homogeneous La$_{0.75}$Sr$_{0.25}$CrO$_3$ polycrystalline sample. Following this, the resultant powder sample was augmented with 5--10 mol\% Cr$_2$O$_3$ \cite{zhu2019method} and mixed before being isostatically pressed into both feed and seed cylindrical rods under a hydrostatic pressure of $\sim$ 70 MPa. These feed and seed rods were then sintered at 1300 $^\circ$C for 36 h in an air environment. To grow single crystals of the La$_{0.75}$Sr$_{0.25}$Cr$_{0.90}$O$_{3-\delta}$ compound, we employed the floating-zone (FZ) technique \cite{li2008synthesis,li2009crystal} using a laser-diode FZ furnace (Model: LZ-FZ-5-200 W-VPO-PC-UM), equipped at the University of Macau in Macao, China \cite{zhu2019method}.

We commenced by grinding a piece of the as-grown La$_{0.75}$Sr$_{0.25}$Cr$_{0.90}$O$_{3-\delta}$ single crystal into fine powder. Subsequently, we conducted sample characterization through X-ray powder diffraction (XRPD) within a range of 2$\theta$ = 10 to 90$^{\circ}$, with a step size of 0.02$^{\circ}$, utilizing an in-house diffractometer (Rigaku, SmartLab 9 kW). The X-ray radiation employed was Cu $K_{\alpha1}$ with a wavelength of 1.54056 {\AA} and $K_{\alpha2}$ with a wavelength of 1.54439 {\AA}, maintaining an intensity ($I$) ratio of $I_{K\alpha2} /$$I_{K\alpha1} =$ 0.5. The scattering configuration adhered to a Bragg-Brentano geometry, with an operating voltage of 45 kV and an operating current of 200 mA. These XRPD measurements took place under ambient conditions. For data refinement, we employed the FULLPROF SUITE software \cite{rodriguez1993recent}. A linear interpolation method was used to establish the background contribution by connecting automatically selected data points. The pseudo-voigt function was chosen to model the shape of the Bragg peaks. We refined a multitude of parameters, including the scale factor, zero shift, peak shape parameters, asymmetry, lattice parameters, atomic positions, and isotropic thermal parameters. In the final stage of analysis, all parameters were collectively refined.

We conducted scanning electron microscope measurements and energy-dispersive X-ray chemical composition analysis on a randomly selected La$_{0.75}$Sr$_{0.25}$Cr$_{0.90}$O$_{3-\delta}$ single crystal using the ZEISS Sigma microscope. To ensure electrical conductivity, the crystal was sputter-coated with gold. The measurements were carried out under a vacuum at an accelerating voltage of 15 kV.

X-ray Laue patterns were obtained from a La$_{0.75}$Sr$_{0.25}$Cr$_{0.90}$O$_{3-\delta}$ single crystal using the X-ray backscattering diffractometer located at Zhuhai UM Science {\&} Technology Research Institute in Zhuhai, China. The collected Laue patterns were simulated using the OrientExpress software \cite{ouladdiaf2006orientexpress}.

Magnetic properties of a La$_{0.75}$Sr$_{0.25}$Cr$_{0.90}$O$_{3-\delta}$ single crystal were characterized using our quantum design physical property measurement system (PPMS DynaCool instrument). Direct current (dc) magnetization measurements were conducted in the temperature range of 1.8 to 400 K under an applied magnetic field of $B$ = 0.1 T, and high-temperature magnetization (300--400 K) was measured under $B$ = 10 T. Two measuring modes were employed: one without the applied magnetic field (ZFC mode) and another under the applied magnetic field (FC mode). Magnetic hysteresis loops were measured in the range of -14 to 14 T at temperatures of 1.8 K and 120 K.

To examine the Cr chemical properties on the surface of a shining \\ La$_{0.75}$Sr$_{0.25}$Cr$_{0.90}$O$_{3-\delta}$ single crystal, we employed high-resolution X-ray photoelectron spectroscopy (Thermo Fisher Scientific, ESCALAB 250Xi). A full-scan spectrum was acquired with an incident energy of 50.0 eV, and a narrow-scan spectrum was obtained with an incident energy of 20.0 eV, both with an energy step size of 0.1 eV.

We acquired the room-temperature light absorption spectrum of a \\ La$_{0.75}$Sr$_{0.25}$Cr$_{0.90}$O$_{3-\delta}$ single crystal utilizing an ultraviolet (UV)-visible-near-infrared (NIR) spectrophotometer (model: Jasco V-770) equipped with an integrating sphere at the University of Macau in Macao, China. The investigation spanned the UV to NIR spectral region, covering an energy range from 0.5 to 5.0 eV. This dataset facilitated the extraction of the material's electrical band gap.

\section{Results and discussion}

\subsection{Single crystal growth}

We optimized the crystal growth parameters by experimenting with various gases, including pure Ar and O$_2$, as well as their mixtures in different rations. Ultimately, we determined that a pure O$_2$ atmosphere (99.999{\%}) at $\sim$ 5--8 atm pressure is the most suitable for stabilizing the floating zone. The feed and seed rods rotate in opposite directions, with speeds ranging from 15 to 30 rpm, while the growth speed falls within the range of 10 to 22 mm{/}h. You can see an example of the as-grown La$_{0.75}$Sr$_{0.25}$Cr$_{0.90}$O$_{3-\delta}$ single crystal in Fig.~\ref{Laue}(a).

\subsection{X-ray Laue study}

\begin{figure*} [t]
\centering
\includegraphics[width = 0.64\textwidth] {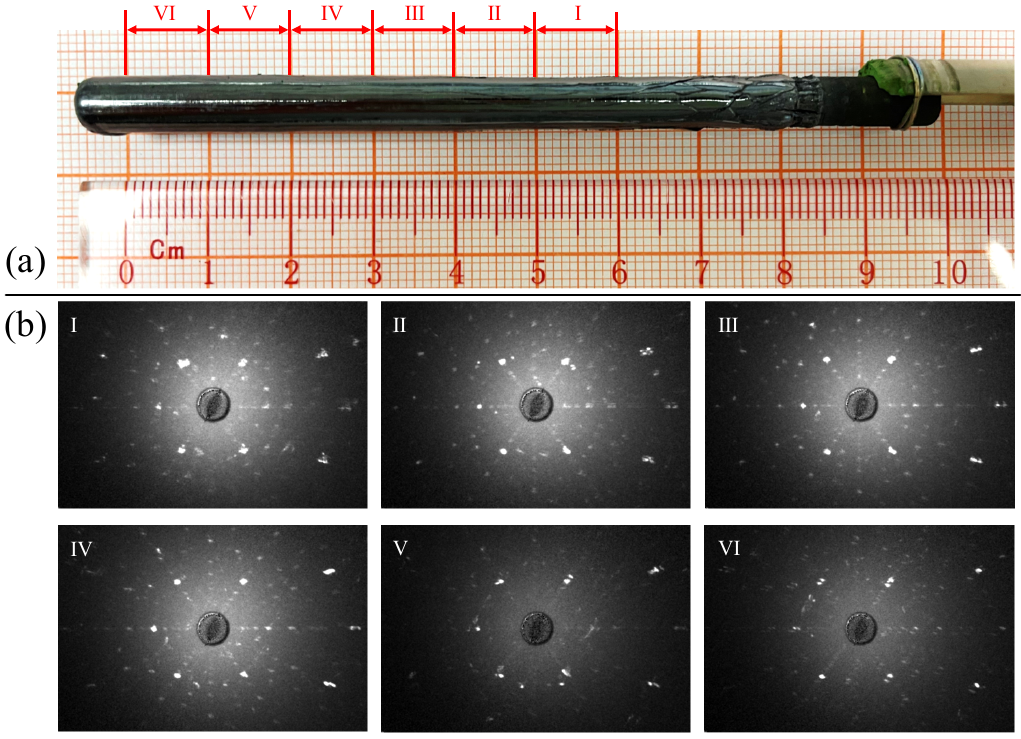}
\caption{(a) A representative image of a La$_{0.75}$Sr$_{0.25}$Cr$_{0.90}$O$_{3-\delta}$ single crystal grown using the laser-diode floating-zone technique. (b) In-house X-ray Laue diffraction patterns of the La$_{0.75}$Sr$_{0.25}$Cr$_{0.90}$O$_{3-\delta}$ single crystal, captured on one side of the crystal at six distinct regions (I, II, III, IV, V, and VI) as indicated in (a) along the growth direction. The discussion regarding the crystal quality in these six regions (\uppercase\expandafter{\romannumeral1} to \uppercase\expandafter{\romannumeral6}) is presented in the text.}
\label{Laue}
\end{figure*}

To confirm the single-crystalline nature of the grown crystal, we conducted X-ray Laue backscattering analysis on a piece of La$_{0.75}$Sr$_{0.25}$Cr$_{0.90}$O$_{3-\delta}$ single crystal (Fig.~\ref{Laue}(a)). We selected six regions, labeled as I to VI, along the crystal growth direction, as indicated in Fig.~\ref{Laue}(a), to assess the crystal quality using X-ray Laue backscattering. The corresponding Laue patterns are displayed in Fig.~\ref{Laue}(b). This investigation revealed that the initial section ($\sim$ 25 mm to the right of region I) exhibits poor quality. This can be attributed to the use of a polycrystalline seed rod, which prolongs the single crystal growth process.

As shown in Fig.~\ref{Laue}(b), we observed nearly identical Laue patterns in all six regions (I to VI) spanning a length of $\sim$ 60 mm along the growth direction. Significantly, these six Laue patterns demonstrate high symmetry, with sets of Laue spots forming lines around the centers of their respective patterns. This symmetry indicates good crystalline quality along the crystal's growth direction. Moreover, the cross-section of the crystal features large grains radiating outward from the center, resembling the shape of a car wheel. This growth pattern differs from the previously reported translational stacking symmetry of grains in rare-earth orthochromate single crystals \cite{zhu2022crystal}. It is possible that the Sr doping induces a metastable structural state in the crystalline structure, allowing for easy transformation between the trigonal ($R\bar{3}c$) and orthorhombic ($Pbnm$) structures, which aligns with our subsequent XRPD studies. The use of high-pressure O$_2$ as the working gas during the crystal growth process may act as a catalyst for inducing this metastable structural state by altering the local crystalline environment of Cr.

\subsection{Structural study}

\begin{figure*} [t]
\centering
\includegraphics[width = 0.64\textwidth] {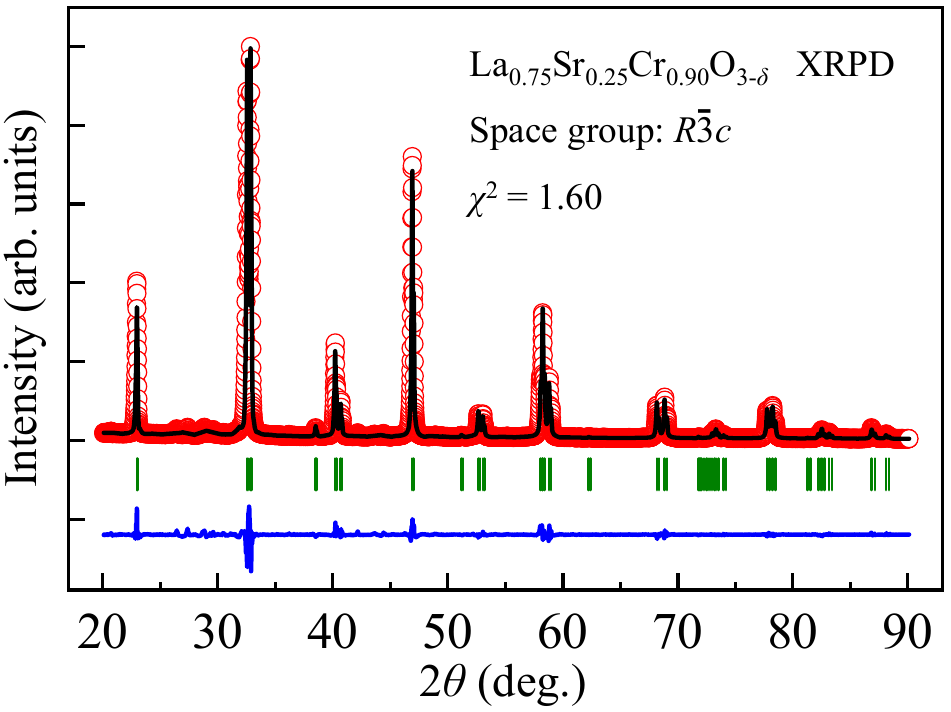}
\caption{Observed (circles) and calculated (solid line) in-house XRPD patterns of a pulverized La$_{0.75}$Sr$_{0.25}$Cr$_{0.90}$O$_{3-\delta}$ single crystal, collected at room temperature. The vertical bars mark the positions of Bragg reflections within the space group $R\bar{3}c$. The bottom curve represents the difference between observed and calculated XRPD patterns. It is worth noting that the minor additional peaks observed at 2$\theta$ = 26.2--32.2$^\circ$ are attributed to radiation contamination resulting from the copper $K_\beta$ wavelength.}
\label{XRPD}
\end{figure*}

\begin{figure*} [t]
\centering
\includegraphics[width = 0.64\textwidth] {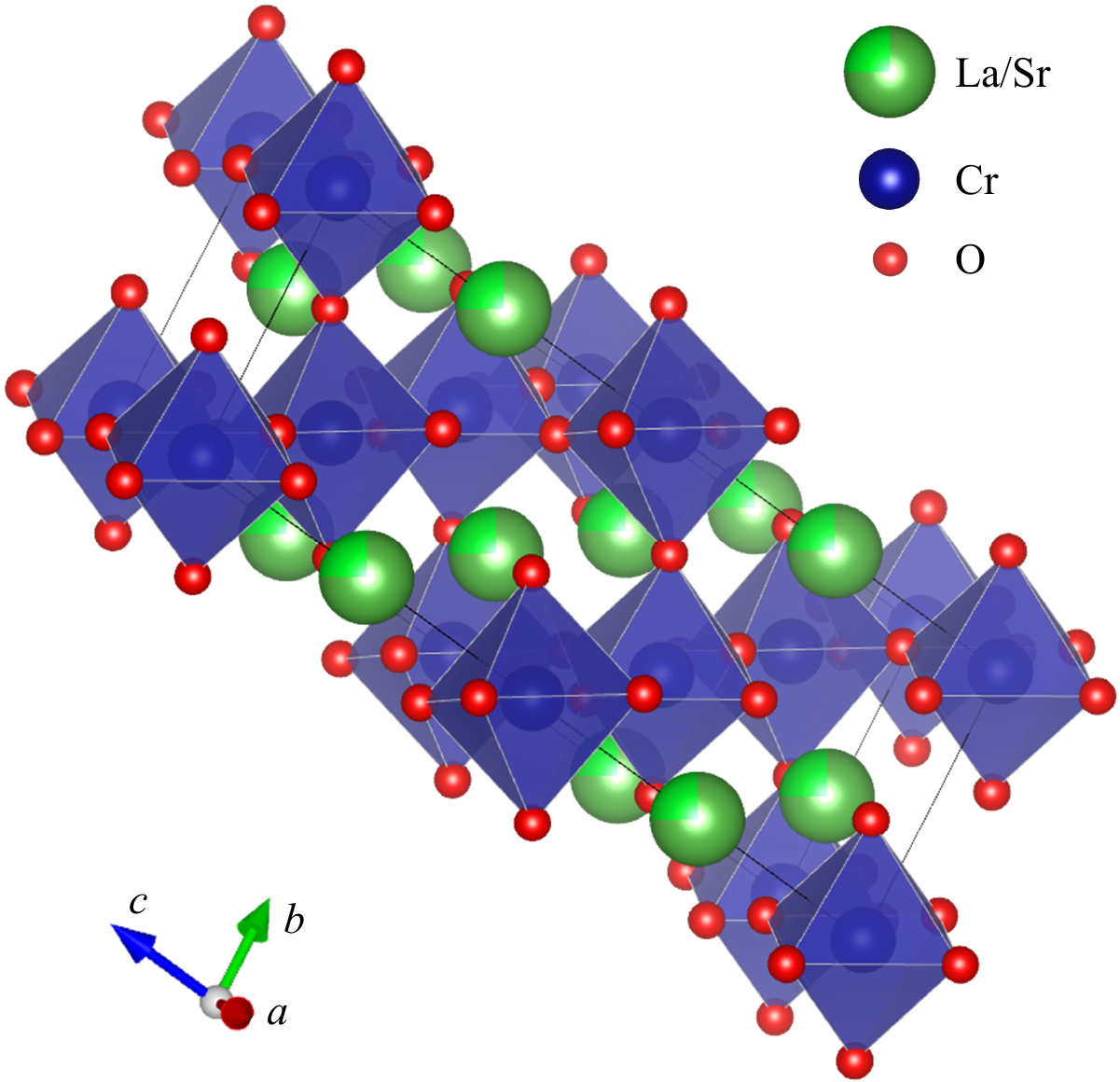}
\caption{Crystal structure (trigonal, space group $R\bar{3}c$) in one unit cell (solid lines) of the single-crystal La$_{0.75}$Sr$_{0.25}$Cr$_{0.90}$O$_{3-\delta}$ compound at room temperature, determined within the present experimental accuracy. The La/Sr, Cr, and O ions are labeled. In this structure model, Cr ions form CrO$_6$ octahedra. It is pointed out that the two oxygen sites are illustrated schematically using the same color code.}
\label{unitcell}
\end{figure*}

To determine the structural phase and extract structure information from the \\ La$_{0.75}$Sr$_{0.25}$Cr$_{0.90}$O$_{3-\delta}$ single crystal, we carried out room-temperature XRPD measurements on a powdered sample of La$_{0.75}$Sr$_{0.25}$Cr$_{0.90}$O$_{3-\delta}$ single crystal. Previous studies have indicated that La$_{0.75}$Sr$_{0.25}$CrO$_3$ exhibits two possible phases at room temperature: either an orthorhombic structure (space group $Pbnm$) \cite{chakraborty2006structural,yusuf2008magnetic} or a trigonal structure (space group $R\bar{3}c$) \cite{khattak1977structural}. We refined the structure of La$_{0.75}$Sr$_{0.25}$Cr$_{0.90}$O$_{3-\delta}$ using the Rietveld refinement method. The collected XRPD pattern and the corresponding refinement results are presented in Fig.~\ref{XRPD}. The observed XRPD pattern can be adequately indexed with either the trigonal or the orthorhombic structure, or a combination of both structures. Although all potential refinements were reasonable in terms of goodness of fit, we ultimately decided to refine the data using only the trigonal structure model within the present experimental accuracy. This decision was made because we did not observe any possible Bragg peak splitting, especially at high 2$\theta$ scattering angles, or any additional Bragg peaks that did not belong to the space group $R\bar{3}c$, indicating a lower level of crystalline symmetry. This finding aligns with heat capacity measurements \cite{matsunaga2008analysis-2}, which shows that when the Sr doping level exceeds $\sim$ 0.13, the lattice structure of La$_{1-x}$Sr$_x$CrO$_3$ tends to become rhombohedral. The extracted room-temperature structural parameters, including lattice constants, unit-cell volume, atomic positions, and isotropic thermal parameters, are listed in Table~\ref{latticep}, with small goodness of fit values indicating a successful Rietveld refinement. The resulting crystal structure is illustrated in Fig.~\ref{unitcell}, where Cr and O ions form CrO$_6$ octahedra, and the La sites are 1/4 occupied by Sr ions. The tilting and rotation of CrO$_6$ octahedra in certain directions may influence the superexchange interactions between Cr $t_{\textrm{2g}}$-$t_{\textrm{2g}}$ orbitals, potentially leading to a canted AFM ordering in the La$_{0.75}$Sr$_{0.25}$CrO$_3$ compound \cite{zhu2022crystal}. Our XRPD study suggests that the structural phase of La$_{0.75}$Sr$_{0.25}$Cr$_{0.90}$O$_{3-\delta}$ compound is strongly influenced by the choice of working gas used in the synthesis precess.

\subsection{Scanning electronic microscopy}

\begin{figure*} [t]
\centering
\includegraphics[width = 0.64\textwidth] {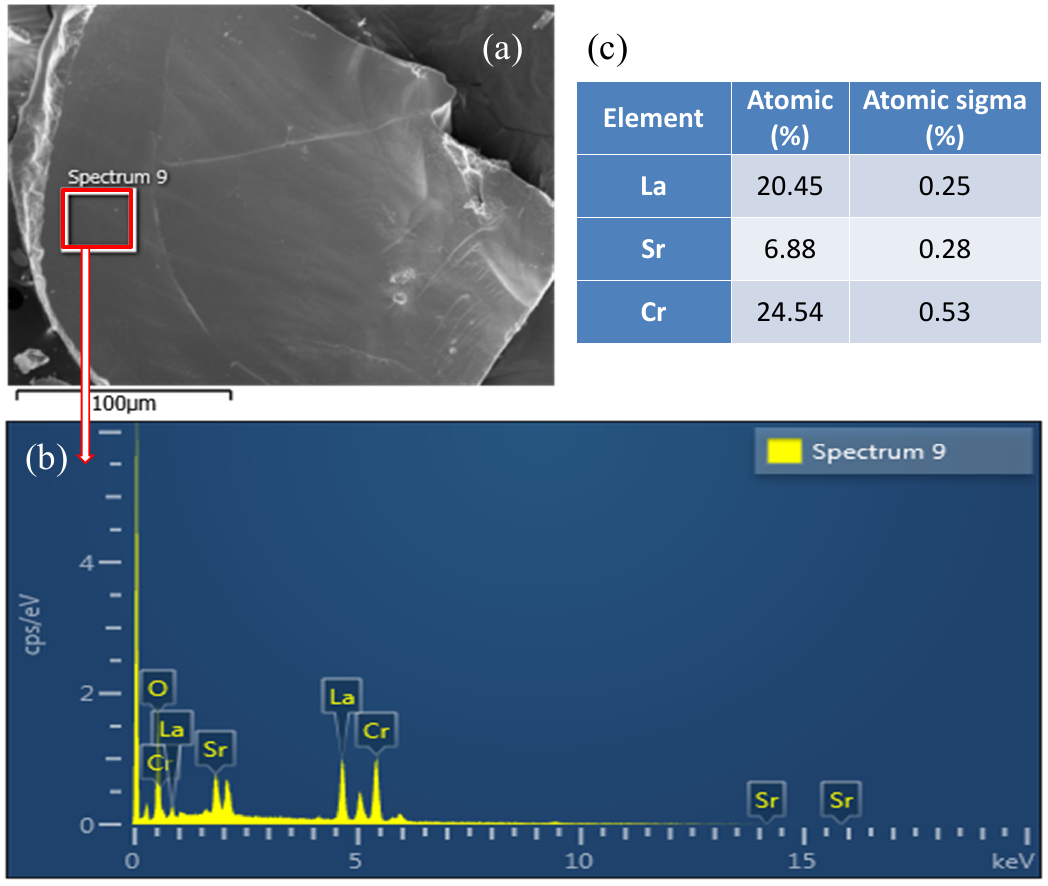}
\caption{(a) Scanning electron microscopy image of a La$_{0.75}$Sr$_{0.25}$Cr$_{0.90}$O$_{3-\delta}$ single crystal. The scale bar represents 100 $\mu$m. (b) Energy-dispersive X-ray chemical composition analysis of the selected area measuring $\sim$ 32 $\times$ 27 $\mu$m (as marked in panel (a)). (c) The extracted atomic percentages ($\%$) of the chemical compositions of La, Sr, and Cr elements along with their respective error bars (atomic sigma). It is pointed out that we excluded the composition analysis of oxygen due to the insensitivity of X-ray detection to the lightweight oxygen element.}
\label{SEM}
\end{figure*}

To analyze the chemical composition of the grown single crystal, we employed scanning electronic microscopy, specifically energy-dispersive X-ray chemical composition analysis. A random section of the La$_{0.75}$Sr$_{0.25}$Cr$_{0.90}$O$_{3-\delta}$ single crystal was selected, and a flat surface measuring $\sim$ 32 $\times$ 27 $\mu$m (the square region marked in Fig.~\ref{SEM}(a)) was chosen for a detailed statistical examination. The corresponding energy-dispersive X-ray spectrum is presently in Fig.~\ref{SEM}(b), and the chemical compositions of the La, Sr, and Cr elements are displayed in Fig.~\ref{SEM}(c) with a ratio of La {:} Sr {:} Cr = 0.75(1) {:} 0.25(1) {:} 0.90(2). Since X-rays do not readily detect oxygen, we cannot determine the content of oxygen in the single crystal. As a result, we determined the chemical formula to be La$_{0.75}$Sr$_{0.25}$Cr$_{0.90}$O$_{3-\delta}$. The presence of Cr vacancies was quite evident and attributed to the significant evaporation of the Cr element during the crystal growth process.

\subsection{Magnetization versus temperature}

\begin{figure*} [t]
\centering
\includegraphics[width = 0.88\textwidth] {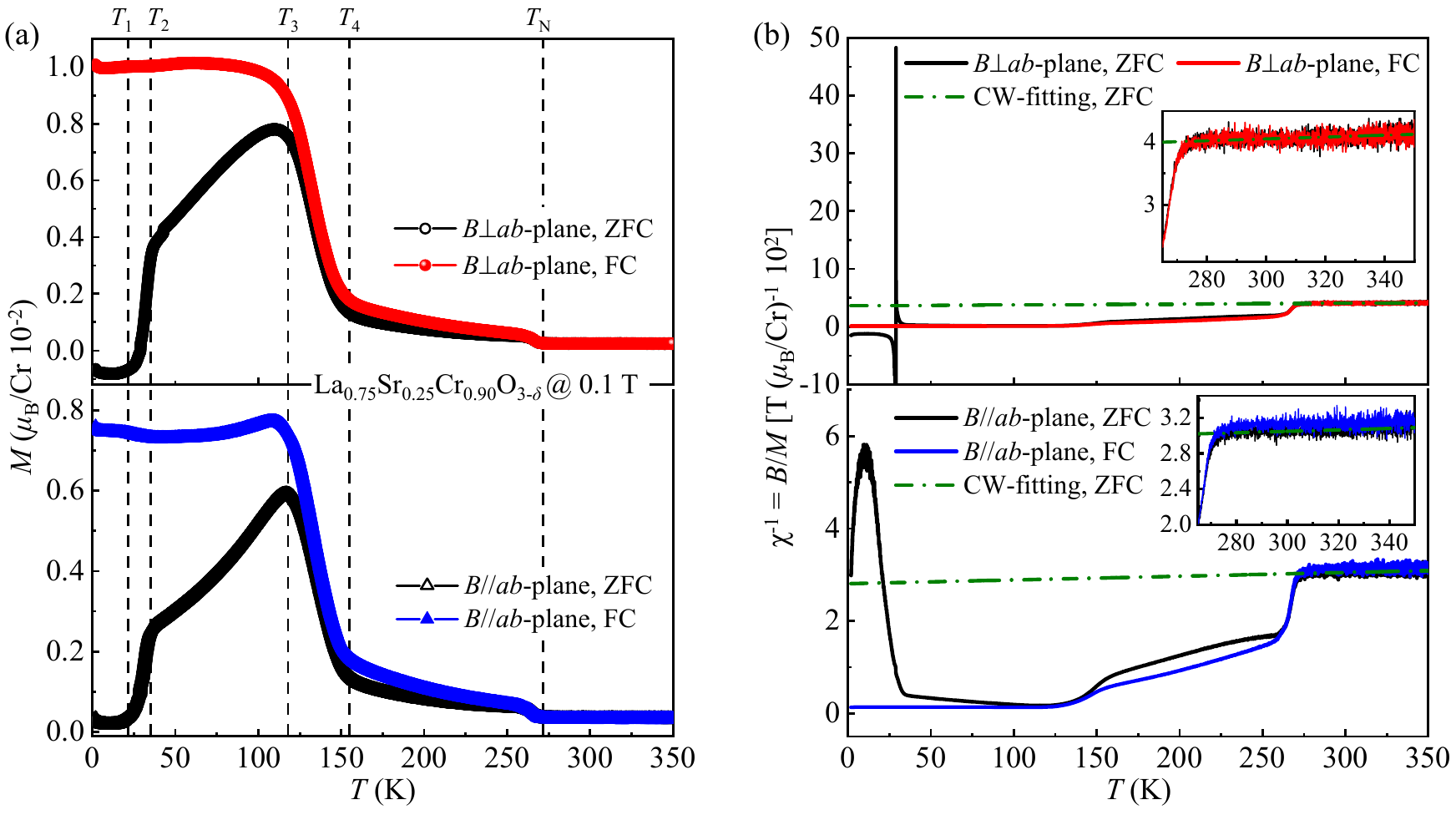}
\caption{(a) ZFC and FC magnetization ($M$) data as a function of temperature, measured when an applied magnetic field $B$ (= 0.1 T) oriented either perpendicular (up panel) or parallel (down panel) to the $ab$-plane of a crystallographically aligned La$_{0.75}$Sr$_{0.25}$Cr$_{0.90}$O$_{3-\delta}$ single crystal. $T_1 =$ 21.50(1) K, $T_2 =$ 34.98(1) K, $T_3 =$ 117.94(1) K, $T_4 =$ 155.01(1) K, and $T_{\textrm{N}} =$ 271.80(1) K represent temperature points marking interesting magnetic anomalies. (b) ZFC and FC inverse magnetic susceptibilities ($\chi^{-1}$) versus temperature of chromium ion in the single-crystal La$_{0.75}$Sr$_{0.25}$Cr$_{0.90}$O$_{3-\delta}$ compound (corresponding to (a)). The dash-dotted line indicates the Curie-Weiss behavior observed in the ZFC data at elevated temperatures between 280 and 350 K. Insets in (b) are enlarged for clarity in displaying the Curie-Weiss fits. It is pointed out that we shortened the vertical axis range in the up panel to enhance visibility of the data changes.}
\label{MT1}
\end{figure*}

Figure~\ref{MT1}(a) shows ZFC and FC magnetization measurements as a function of temperature under an applied magnetic field of $B$ = 0.1 T. The magnetization of the \\ crystallographically-orientated La$_{0.75}$Sr$_{0.25}$Cr$_{0.90}$O$_{3-\delta}$ single crystal shows a paramagnetic (PM) state above $T_\textrm{N}$ = 271.80(1) K. Below $T_\textrm{N}$, when $B$ is perpendicular to the $ab$-plane, the ZFC magnetization displays a small but distinct step increase from $\sim$ 2.57 $\times$ 10$^{-4}$ to $\sim$ 5.06 $\times$ 10$^{-4}$ $\mu_\textrm{B}/$Cr within a temperature range of $\sim$ 11.47 K, indicating a magnetic phase transition (i.e., a canted AFM transition). Subsequently, the magnetization steadily rises with decreasing temperature until $T_4 =$ 155.01(1) K. Below $T_4$, the magnetization increases sharply and reaches near saturation at $T_3 =$ 117.94(1) K. Below $T_3$, the FC magnetization stabilizes, while the ZFC magnetization undergoes a rapid decline, resulting in a considerable deviation from the FC magnetization. The corresponding stable FC magnetic moment per formula unit when $B \perp ab$-plane is $\sim$ 1.04 $\times$ 10$^{-2}$ $\mu_\textrm{B}$/Cr at 1.8 K. The ZFC magnetization between $T_2 =$ 34.98(1) K and $T_1 =$ 21.50(1) K decreases more rapidly than that between $T_2 \leq T \leq T_3$, indicating a possible change in AFM spin directions. Below $T_1$, the ZFC magnetization gets flatted. Consequently, the La$_{0.75}$Sr$_{0.25}$Cr$_{0.90}$O$_{3-\delta}$ single crystal exhibits a series of magnetic anomalies during cooling. The corresponding transition temperatures were confirmed using the inverse magnetic susceptibility ($\chi^{-1}$) of ZFC magnetization with respect to temperature as shown in Fig.~\ref{MT1}(b). Here, we define the magnetic anomaly temperature as the point at which a kink appears in the $\chi^{-1}(T)$ curve.

The bifurcation between ZFC and FC magnetization curves below the irreversible temperature ($T_3$) indicates the possibility of a spin-glassy magnetic state at low temperatures \cite{mazumdar2021structural}. The complexity of magnetic anomalies may arise from a temperature-dependent competition among factors such as Cr spin interactions, crystal field effects, single-ion anisotropy, and Dzyaloshinskii-Moriya interactions \cite{li2016possible, li2014incommensurate}. In the La$_{0.75}$Sr$_{0.25}$CrO$_3$ compound, oxygen atoms form octahedral coordination environments around chromium ions. These local crystalline environments of Cr ions result in variations in electron density distribution and the overall charge of octahedra. This charge redistribution, in conjunction with electron orbital overlap mediated by oxygen atoms, leads to Cr spin superexchange interactions among neighboring chromium ions.

Figure~\ref{MT1}(b) shows the deduced inverse magnetic susceptibility, i.e., $\chi^{-1}$ = $B/M$, in correlation with the data shown in Fig.~\ref{MT1}(a). It is obvious that the inverse magnetic susceptibility exhibits distinct reactions to the magnetic anomalies. The high-temperature inverse magnetic susceptibility in a pure PM state observes the Curie-Weiss (CW) law \cite{li2009crystal, li2007neutron},
\begin{eqnarray}
\chi^{-1}(T) = \frac{T - \theta_{\textrm{CW}}}{C} = \frac{3k_\textrm{B}(T - \theta_{\textrm{CW}})}{N_\textrm{A} \mu^2_{\textrm{eff}}},
\label{CWLaw}
\end{eqnarray}
where $\theta_\textrm{CW}$ is the PM Curie temperature, $C$ is the Curie constant, $k_\textrm{B}$ = 1.38062 $\times$ 10$^{-23}$ J/K is the Boltzmann constant, $N_\textrm{A}$ = 6.022 $\times$ 10$^{23}$ mol$^{-1}$ is the Avogadro's number, and $\mu_{\textrm{eff}}$ = $g \mu_\textrm{B} \sqrt{J(J + 1})$ is the effective PM moment. We fit the high-temperature data from 280 to 350 K with Equ.~\ref{CWLaw}, shown as the dash-dotted lines in Fig.~\ref{MT1}(b), which yields effective PM moments ${\mu}_{\textrm{eff}}$ = 5.7(2) ${\mu}_\textrm{B}$ ($B \perp ab$-plane) and 7.8(3) ${\mu}_\textrm{B}$ ($B \parallel ab$-plane). The extracted ${\mu}_{\textrm{eff}}$ values are considerably larger than the theoretically calculated value, $\mu_{\textrm{eff{\_}theo}} =$ 3.873 $\mu_\textrm{B}$ \cite{zhu2020crystalline}, implying a possible presence of strong spin fluctuations between 280 and 350 K \cite{li2012possible}.

\begin{figure*} [t]
\centering
\includegraphics[width = 0.88\textwidth] {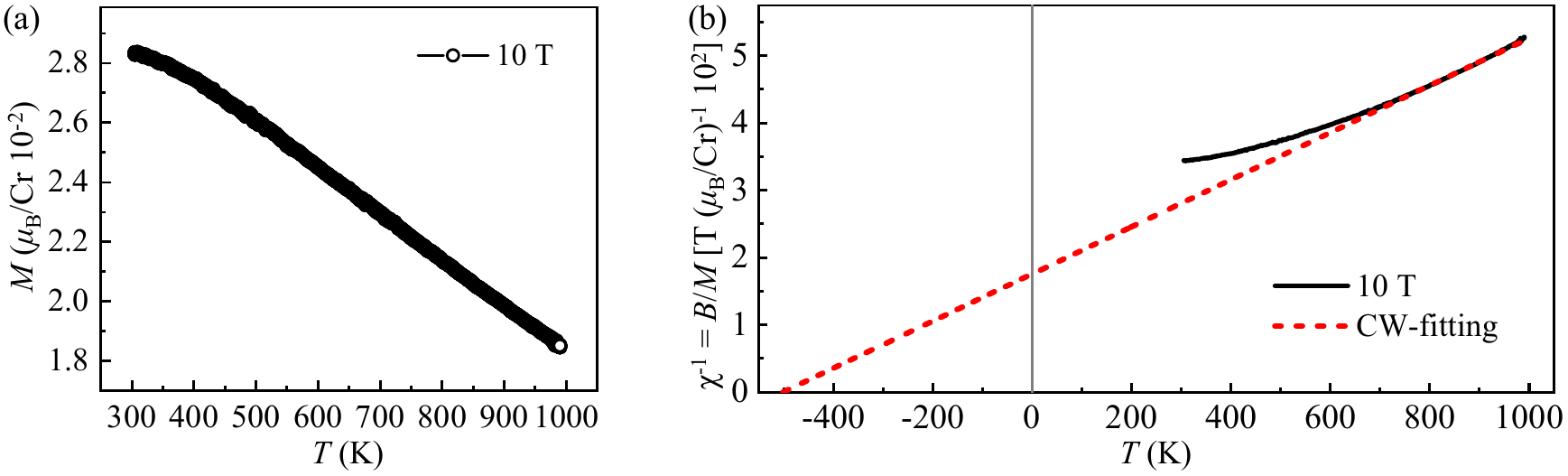}
\caption{(a) Magnetization ($M$) data at high temperatures, ranging from 300 to 1000 K, for a randomly orientated La$_{0.75}$Sr$_{0.25}$Cr$_{0.90}$O$_{3-\delta}$ single crystal. These measurements were taken with an applied magnetic field of $B$ = 10 T. (b) Corresponding inverse magnetic susceptibilities ($\chi^{-1}$) for the chromium ion versus temperature. The dash-dotted line indicates the Curie-Weiss behavior observed in the data between 750 and 950 K.}
\label{MTH}
\end{figure*}

It is well known that the CW law is applicable exclusively to the pure PM state. We extended our magnetization measurements over a temperature range from 300 to 1000 K as shown in Fig.~\ref{MTH}(a). We attempted to fit the data within the temperature range of 750-950 K to the CW law (Fig.~\ref{MTH}(b)). This fitting resulted in an effective PM moment of 3.570(1) $\mu_\textrm{B}$, which is both reasonable and smaller than the corresponding theoretical value of 3.873 $\mu_\textrm{B}$. The resultant PM CW temperatures ${\theta}_\textrm{CW}$ = --501.4(4) K. We also tentatively calculated the magnetic frustration parameter $f = \frac{|\theta_{\textrm{CW}}|}{T_\textrm{N}}$ \cite{zhu2020high} for the single-crystal La$_{0.75}$Sr$_{0.25}$Cr$_{0.90}$O$_{3-\delta}$, which yields an $f_{\textrm{Cr}}$ value of $\sim$ 1.84. This indicates that the Cr spin moments experience mild frustration.

\subsection{Magnetization versus applied magnetic field}

\begin{figure*} [t]
\centering
\includegraphics[width = 0.88\textwidth] {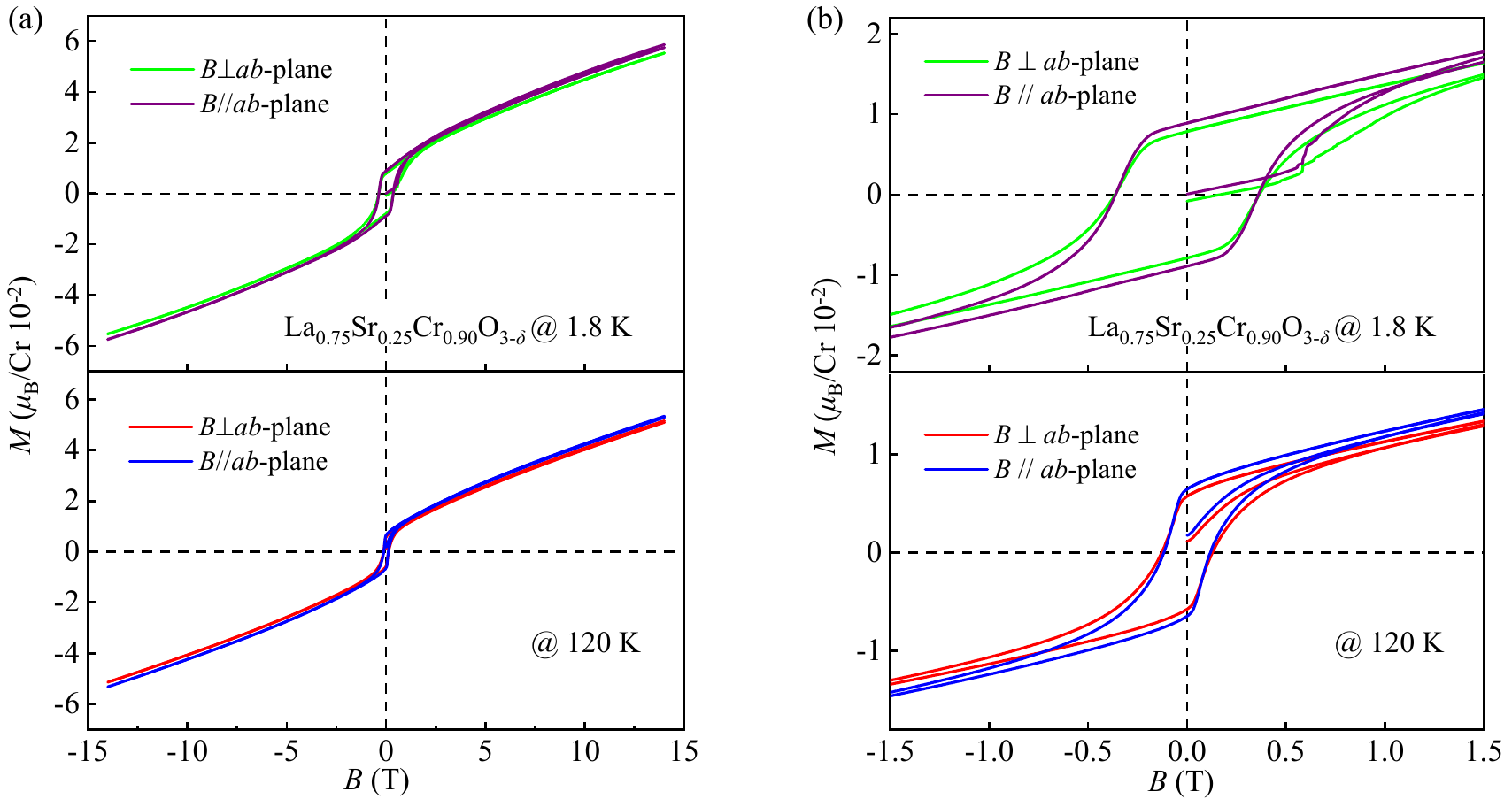}
\caption{(a) Magnetic hysteresis loops recorded for a crystallographically-aligned La$_{0.75}$Sr$_{0.25}$Cr$_{0.90}$O$_{3-\delta}$ single crystal, measured at 1.8 K (up panel) and 120 K (down panel). These measurements were conducted under varying applied magnetic fields ($B$) ranging from -14 to 14 T, with $B$ oriented either perpendicular or parallel to the $ab$-plane. (b) To enhance clarity and highlight the magnetic hysteresis effect observed in (a), we have presented the data only for the range of -1.5 to 1.5 T.}
\label{MH}
\end{figure*}

Figure~\ref{MH} shows the ZFC magnetization measurements as a function of the applied magnetic field, i.e., $M(B)$, at two temperature points of 1.8 K (up panels) and 120 K (down panels), both below $T_\textrm{N}$. At both of these temperatures, the $M(B)$ curves exhibit well-defined magnetic hysteresis loops, regardless of the orientation of the applied magnetic field. This observation indicates the presence of a ferromagnetic component contributing to the measured magnetization.

As the applied magnetic field ($B$) increases, the measured magnetization exhibits nearly linear growth and does not reach saturation, even when exposed to a field as high as 14 T. When the applied magnetic field is aligned with the $ab$-plane ($B \parallel ab$-plane), the measured magnetization at 1.8 K and 14 T equals $\sim$ 5.86 $\times$ 10$^{-2}$ $\mu_\textrm{B}/$Cr. This value is significantly smaller (only $\sim$ 1.95{\%}) than the theoretical saturation value of 3 $\mu_\textrm{B}$ of ionic Cr$^{3+}$ ions \cite{zhu2020crystalline}.

Figure~\ref{MH}(b) shows a weak anisotropic effect of the magnetic hysteresis loops. The width of the hysteresis loop at 1.8 K (up panel) is obviously larger than that at 120 K (down panel). At 1.8 K, the coercivity $H_c$ is measured to be 3622(4) Oe when \emph{B}$\parallel$\emph{ab}-plane and 3652(3) Oe when \emph{B}$\perp$\emph{ab}-plane. At 120 K, the coercivity is recorded as $H_c =$ 1185(7) Oe when \emph{B}$\parallel$\emph{ab}-plane and $H_c =$ 1290(8) Oe when \emph{B}$\perp$\emph{ab}-plane.

\subsection{X-ray photoelectron spectroscopy}

\begin{figure*} [t]
\centering
\includegraphics[width = 0.64\textwidth] {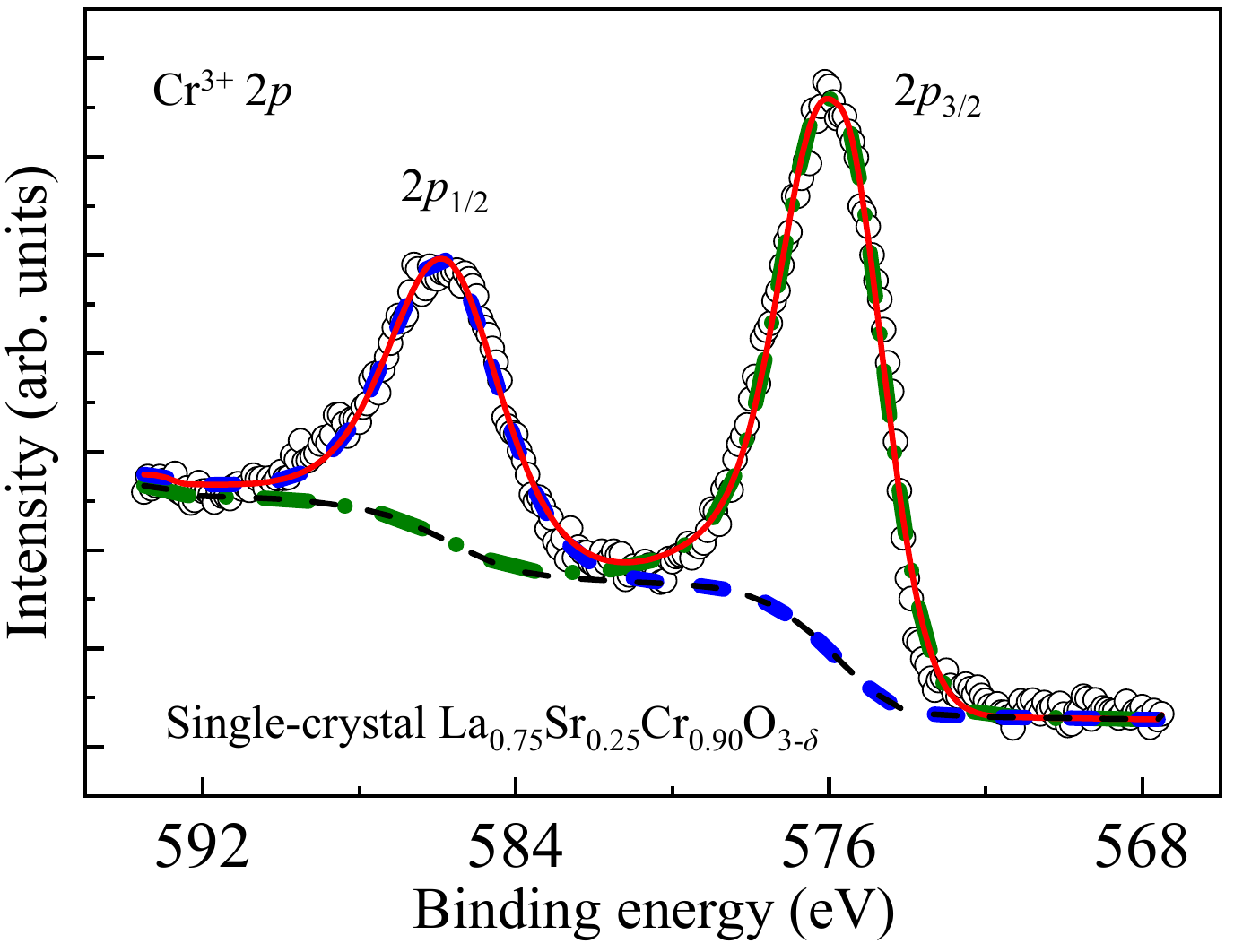}
\caption{X-ray photoelectron spectroscopy spectrum (circles) captures the Cr$^{3+}$ 2$p$-edge at the surface of a La$_{0.75}$Sr$_{0.25}$Cr$_{0.90}$O$_{3-\delta}$ single crystal. The spectrum reveals distinct features: the Cr$^{3+}$ $2p_{1/2}$ peak (on the short-dashed blue line) at 585.9 eV and the Cr$^{3+}$ $2p_{3/2}$ peak (on the short-dash-dotted olive line) at 576 eV. The solid red line indicates the fitted curve, which corresponds to the intensity sum of the Cr$^{3+}$ $2p_{1/2}$ and Cr$^{3+}$ $2p_{3/2}$ spectra. Meanwhile, the dashed black line denotes the contribution from the background.}
\label{XPS}
\end{figure*}

Different oxidation states of chromium are associated with distinct X-ray photoelectron spectroscopy spectra, characterized by chemical shifts or changes in binding energy. In Fig.~\ref{XPS}, the Cr 2$p$ X-ray photoelectron spectroscopy of the surface of a single-crystal La$_{0.75}$Sr$_{0.25}$Cr$_{0.90}$O$_{3-\delta}$ sample is displayed. According to the X-ray photoelectron spectroscopy handbook \cite{wagner1979handbook}, a meticulous quantitative analysis of the spectrum reveals that the peaks at 585.9 eV and 576 eV correspond to Cr$^{3+}$ $2p_{1/2}$ and Cr$^{3+}$ $2p_{3/2}$, respectively. At the curent resolution, there is no observable evidence of metallic Cr, Cr$^{4+}$ and Cr$^{6+}$ states within the La$_{0.75}$Sr$_{0.25}$Cr$_{0.90}$O$_{3-\delta}$ single crystal. Thus, the observed Cr spectrum is consistent solely with the Cr$^{3+}$ oxidation state.

\subsection{Light absorption}

\begin{figure*} [t]
\centering
\includegraphics[width = 0.64\textwidth] {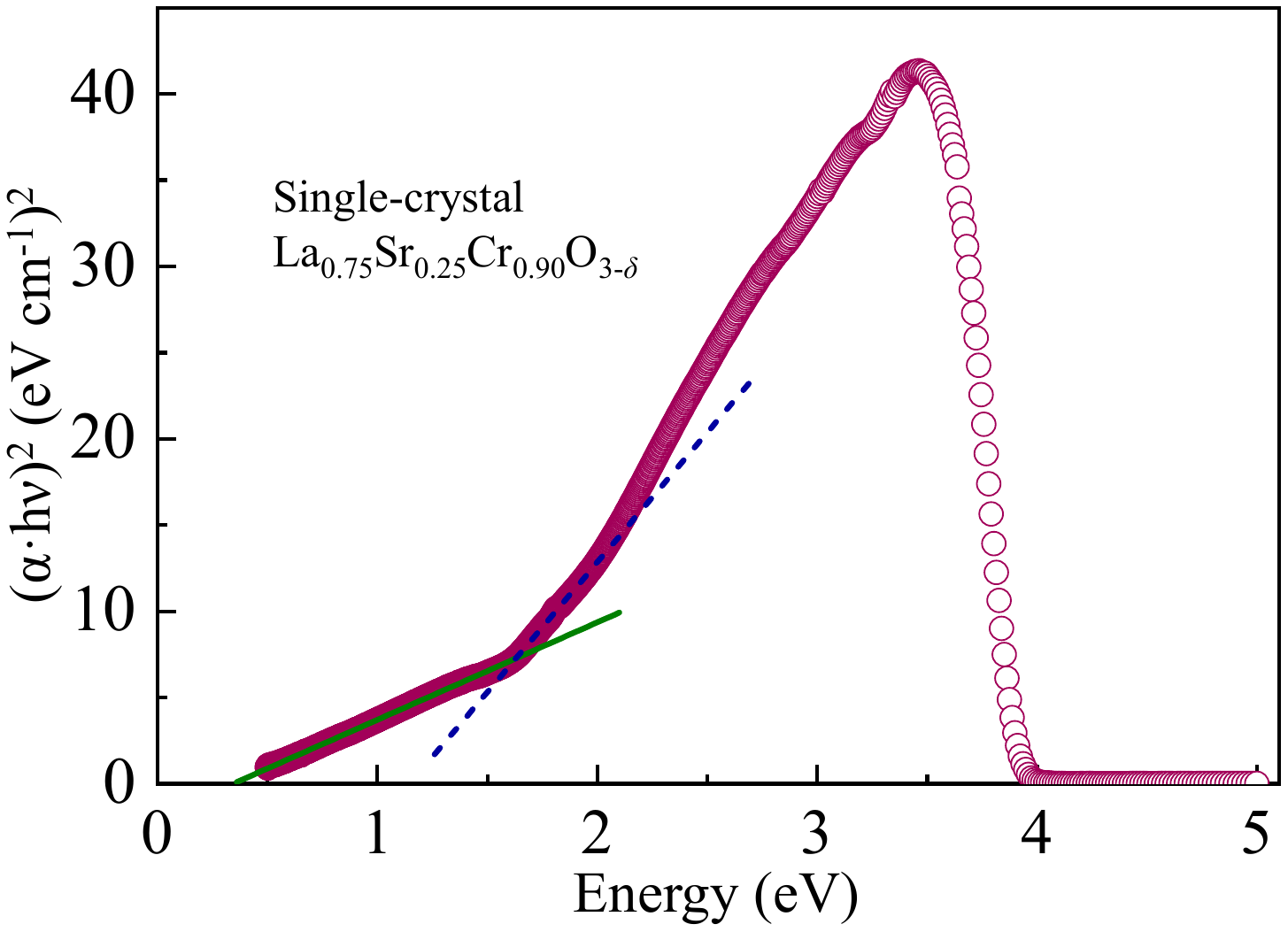}
\caption{Tauc plot for the La$_{0.75}$Sr$_{0.25}$Cr$_{0.90}$O$_{3-\delta}$ single crystal. Linear fits (solid and short-dashed lines) have been applied to the data within the energy ranges of 0.52--1.55 eV and 1.63--2.00 eV, respectively. Extrapolations were performed for both fits.}
\label{TPlot}
\end{figure*}

We utilized the Tauc-plot method \cite{Patrycja2018} to experimentally determine the band gap energy of the La$_{0.75}$Sr$_{0.25}$Cr$_{0.90}$O$_{3-\delta}$ single crystal, that is,
\begin{equation}
(\alpha \cdot h \upsilon)^{\frac{1}{n}} = C (h \upsilon - E_\textrm{g}),
\label{BGEC}
\end{equation}
where $\alpha$ represents the energy-dependent absorption coefficient, \emph{h} is the Planck constant ($h = 6.626 \times 10^{-34}$ J$\cdot$s), $\upsilon$ is the photon's frequency, $n = \frac{1}{2}$ for the direct transition band gap, $C$ is a constant, and $E_\textrm{g}$ denotes the band gap energy. The Tauc plot for the La$_{0.75}$Sr$_{0.25}$Cr$_{0.90}$O$_{3-\delta}$ single crystal is depicted in Fig.~\ref{TPlot}, displaying a pronounced increase in absorption with rising energy at around 1.6 eV, a characteristic feature of semiconductors. We performed linear fits on the two distinct portions of the data (Fig.~\ref{TPlot}), leading to an estimate for the band gap width. This value is determined by the energy axis intersection point of the two linear fits, denoted as $E_\textrm{g} =$ 1.630(8) eV.

\subsection{Possible origins of magnetic anomalies}

\begin{figure*} [t]
\centering
\includegraphics[width = 0.64\textwidth] {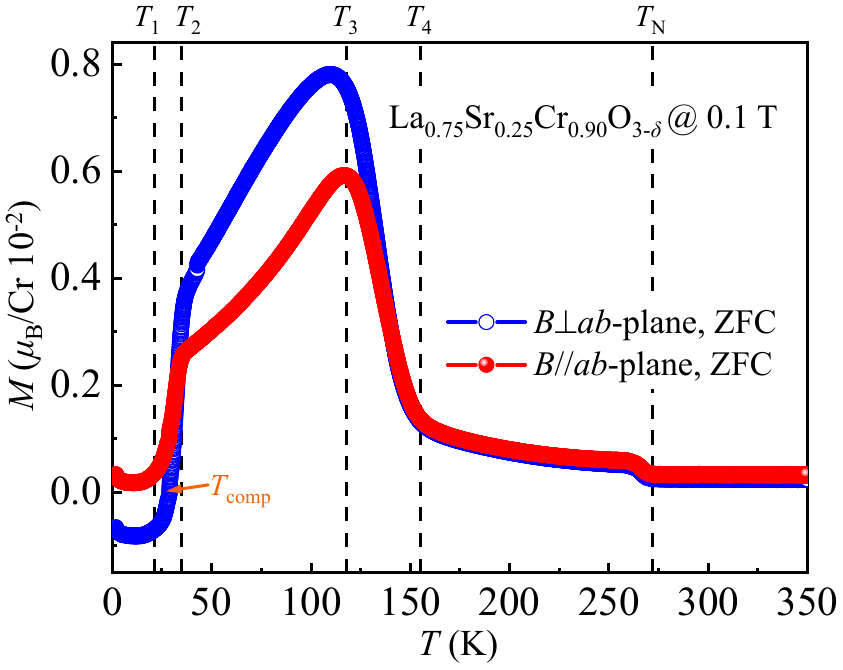}
\caption{ZFC magnetization ($M$) data as a function of temperature, measured under an applied magnetic field ($B$) of 0.1 T. These measurements were taken with $B$ oriented both perpendicular and parallel to the $ab$-plane of a crystallographically-aligned La$_{0.75}$Sr$_{0.25}$Cr$_{0.90}$O$_{3-\delta}$ single crystal. $T_1 =$ 21.50(1) K, $T_2 =$ 34.98(1) K, $T_3 =$ 117.94(1) K, $T_4 =$ 155.01(1) K, and $T_{\textrm{N}} =$ 271.80(1) K mark magnetic anomaly temperatures in the data.}
\label{MT2}
\end{figure*}

Based on our magnetization study, we have identified five critical temperature points ($T_1 =$ 21.50(1) K, $T_2 =$ 34.98(1) K, $T_3 =$ 117.94(1) K, $T_4 =$ 155.01(1) K, and $T_{\textrm{N}} =$ 271.80(1) K) as listed in Table~\ref{Comp}. To investigate the magnetic anomalies and their origins, we conducted a schematic comparison of ZFC magnetization measurements when applied magnetic fields are oriented either perpendicular or parallel to the \emph{ab}-plane (see Fig.~\ref{MT2}). At $T_{\textrm{N}} =$ 271.80(1) K, the ZFC magnetization exhibits a small step increase, suggestive of a transition that resembles the formation of a ferromagnetic state. However, it is worth noting that the measured magnetization at this point is only $\sim$ 1.95{\%} (observed at 1.8 K and 14 T when $B \parallel ab$-plane, as seen in Fig.~\ref{MH}) of the theoretical saturation value. Hence, we interpret the phase transition at $T_{\textrm{N}}$ as a canted AFM transition for the single-crystal La$_{0.75}$Sr$_{0.25}$Cr$_{0.90}$O$_{3-\delta}$. In accordance with studies \cite{tezuka1998magnetic,tezuka1998magnetic-2,nakamura2005analysis,matsunaga2008analysis} conducted on polycrystalline La$_{0.85}$Sr$_{0.15}$CrO$_3$ samples, we may attribute the phase transition at $T_4$ to a structural phase transition, likely from trigonal to orthorhombic, and/or to a spin reorientation transition. At $T_3$, the thermally aligned Cr spins begin to become antiparallel, leading to a reduction in ZFC magnetization and the establishment of an AFM1 state. In the vicinity of $T_2$, $M(\emph{B}\perp$ \emph{ab}) remains larger than $M(\emph{B}\parallel$ \emph{ab}), while around $T_1$, $M(\emph{B}\perp$ \emph{ab}) becomes smaller than $M(\emph{B}\parallel$ \emph{ab}), suggesting a potential spin-reorientation transition, possibly involving the transfer of the AFM axis from the $ab$-plane to the \emph{c} axis or the further development of a ferromagnetic component along the $c$ axis. Below $T_1$, $M(\emph{B}\parallel$ \emph{ab}) $>$ $M(\emph{B}\perp$ \emph{ab}), indicating the establishment of a final magnetic state (AFM2) (Table~\ref{Comp}).

The origin of the AFM canting and spin reorientation can likely be attributed to anisotropic spin exchange interactions. Below $T_\textrm{N}$, the CrO$_6$ octahedra experience distortion in La$_{0.75}$Sr$_{0.25}$CrO$_3$ compound, causing the Cr-O-Cr bond angles to deviate from 180$^\circ$. This deviation results in an imperfect superexchange interaction, which may include possible Dzyaloshinskii-Moriya antisymmetric exchange interactions. Consequently, a canted AFM magnetic ordering emerges within the La$_{0.75}$Sr$_{0.25}$CrO$_3$ compound. This magnetic ordering introduces a weak ferromagnetic component, with Cr spins deviating from the AFM axis as temperature decreases.

In this study, the observed canted AFM transition temperature ($T_{\textrm{N}} =$ 271.80(1) K) in the La$_{0.75}$Sr$_{0.25}$Cr$_{0.90}$O$_{3-\delta}$ single crystal is notably lower than that of polycrystalline LaCrO$_3$ ($T_{\textrm{N}} =$ 295 K) and La$_{0.9}$Sr$_{0.1}$CrO$_3$ ($T_{\textrm{N}} =$ 292 K) samples \cite{SILVAJR2022139278}. Our SEM study confirms deficiencies in both Cr and oxygen sites. Interestingly, our XPS study rules out the presence of Cr$^{4+}$ and Cr$^{6+}$ oxidation states. It is well-known that superexchange interactions between neighboring spins of Cr$^{3+}$ ions are realized through a virtual charge transfer via the bridge of O$^{2-}$ ions. Hence, the reduction in $T_{\textrm{N}}$ in the single-crystal La$_{0.75}$Sr$_{0.25}$Cr$_{0.90}$O$_{3-\delta}$ primarily stems from the diminished degree of $t$-$e$ orbital hybridization in the Cr$^{3+}$($t_\texttt{2g$\uparrow$}^3$)-O$^{2-}$-Cr$^{3+}$($e_\texttt{g}^0$) interactions \cite{zhu2022crystal}, attributed to significant Cr and O vacancies.

\subsection{Negative magnetization behavior}

Negative magnetization (NM) behavior was observed in Pr$_{1-x}$Nd$_x$MnO$_3$ ($x$ = 0.3 and 0.5) \cite{biswas2014magnetization}, where the FC dc magnetization under the applied magnetic field of 50 Oe undergoes a positive-to-negative magnetization transition around 18 and 22 K, respectively. Similar NM observations were also observed in La$_{0.75}$Nd$_{0.25}${CrO}$_3$ \cite{Khomchenko2008} and RETMO$_3$ (RE = rare earth, TM = Cr/Mn) \cite{biswas2018negative} perovskites. In these samples, the NM is mainly due to the negative exchange coupling between the canted-AFM and FM sublattices.

The ZFC NM observed in orthochromates \cite{kumar2015phenomenon, yoshii2001magnetic, jaiswal2010magnetic, cao2014temperature} is an example of a unique case, where the NM is due to fact that the canted AFM spins inducing an internal magnetic field stronger than the applied magnetic field \cite{kumar2015phenomenon}.

It is crucial to emphasize that the La$_{0.75}$Sr$_{0.25}$CrO$_3$ single crystal demonstrates canted antiferromagnetism, specifically correlated to 3$d$ Cr ions. Magnetization measurements were performed using a Quantum Design PPMS DynaCool, with the applied magnetic field consistently oriented vertically upward. When magnetization is performed in the vertically downward direction in the single-crystal La$_{0.75}$Sr$_{0.25}$CrO$_3$, the output signal naturally registers as negative. This study reveals an interesting observation: at a compensation (comp) temperature ($T_{\textrm{comp}}$) of approximately 28.90 K, defined as the thermal point where the measured magnetization transforms from positive to negative values, the ZFC magnetization when $B$(= 0.1 T) $\perp$ \emph{ab}-plane enters the negative value regime (Fig.~\ref{MT2}). One plausible explanation for the observed $M(\emph{B}\perp$ \emph{ab}) $<$ 0 behavior is that the FM component of the canted spins becomes downward.

Consequently, the AFM2 state exhibits a more complex canting configuration, a phenomenon that could be elucidated through further studies of the detailed crystal and magnetic structures with temperature, obtained through elastic neutron-scattering.

\section{Conclusions}

To summarize, we have successfully paved the way for growing large and good-quality La$_{0.75}$Sr$_{0.25}$Cr$_{0.90}$O$_{3-\delta}$ single crystals. Our optimization of growth parameters, such as the use of a pure O$_2$ working environment and a growth speed of 10--22 mm{/}h, has been instrumental in achieving this. However, the high evaporation of Cr during the crystal growth process has led to a significant Cr deficiency in the La$_{0.75}$Sr$_{0.25}$Cr$_{0.90}$O$_{3-\delta}$ single crystals. For instance, the actual formula was determined to be La$_{0.75}$Sr$_{0.25}$Cr$_{0.90}$O$_{1-\delta}$. Our XRPD study reveals a single room-temperature phase for the La$_{0.75}$Sr$_{0.25}$Cr$_{0.90}$O$_{3-\delta}$ single crystals, characterized by a trigonal structure with the space group $R\bar{3}c$. This finding signifies the good crystalline quality of the crystals. Additionally, X-ray photoelectron spectroscopy measurements have shown that all Cr ions in La$_{0.75}$Sr$_{0.25}$Cr$_{0.90}$O$_{3-\delta}$ single crystals exhibit a consistent oxidation state of Cr$^{3+}$. Upon cooling, the La$_{0.75}$Sr$_{0.25}$Cr$_{0.90}$O$_{3-\delta}$ single crystal, characterized by a band gap of 1.630(8) eV, undergoes a canted AFM transition at $T_{\textrm{N}} =$ 271.80(1) K, followed by a possible structural and/or spin-reorientation transition at $T_4 =$ 155.01(1) K. Further cooling results in the alignment of Cr spins, which begins to form an AFM arrangement with long-range order at $T_3 =$ 117.94(1) K. At $T_2 =$ 34.98(1) K, the AFM spins may experience further canting or the ferromagnetic component may rotate, leading to a change in magnetic anisotropy, or a second spin-reorientation transition occurs. Further experimental studies on these grown single crystals would be of great interest to deepen our understanding.

\section*{\emph{CRediT authorship contribution statement}}

\textbf{Kaitong Sun:} Conceptualization, Methodology, Software, Investigation, Writing - Original Draft, Visualization.
\textbf{Yinghao Zhu:} Conceptualization, Methodology, Software, Investigation, Visualization.
\textbf{S. Yano:} Conceptualization, Methodology, Investigation, Supervision, Visualization.
\textbf{Qian Zhao:} Methodology, Software, Investigation, Visualization.
\textbf{Muqing Su:} Methodology, Software, Investigation, Visualization.
\textbf{Guanping Xu:} Methodology, Software, Investigation, Visualization.
\textbf{Ruifeng Zheng:} Methodology, Software, Investigation.
\textbf{Ying Ellie Fu:} Methodology, Software, Investigation, Visualization.
\textbf{Hai-Feng Li:} Conceptualization, Validation, Writing - Review \& Editing, Visualization, Supervision, Project administration, Funding acquisition.

\section*{Declaration of Competing Interest}

The authors declare that they have no known competing financial interests or personal relationships that could have appeared to influence the work reported in this paper.

\section*{Data availability}

Data will be made available on request.

\section*{Acknowledgments}

S.Y. is financially supported by the National Science and Technology Council, Taiwan, with Grants No. 110-2112-M-213-013 and No. 111-2112-M-213-023. The work at the University of Macau was supported by the Science and Technology Development Fund, Macao SAR (File Nos. 0090{/}2021{/}A2 and 0049{/}2021{/}AGJ), University of Macau (MYRG2020{-}00278{-}IAPME), and the Guangdong{-}Hong Kong{-}Macao Joint Laboratory for Neutron Scattering Science and Technology (Grant No. 2019B121205003).

\bibliographystyle{elsarticle-num}
\bibliography{LSGOSG}

\begin{thebibliography}{10}
\expandafter\ifx\csname url\endcsname\relax
  \def\url#1{\texttt{#1}}\fi
\expandafter\ifx\csname urlprefix\endcsname\relax\def\urlprefix{URL }\fi
\expandafter\ifx\csname href\endcsname\relax
  \def\href#1#2{#2} \def\path#1{#1}\fi

\bibitem{tan2013catalytic}
W.~Tan, Q.~Zhong, D.~Xu, H.~Yan, X.~Zhu, {Catalytic activity and sulfur
  tolerance for {Mn}-{substituted} {La}$_{0.75}${Sr}$_{0.25}${CrO}$_{3 \pm
  \delta}$ in gas containing {H}$_2$S}, International Journal of Hydrogen
  Energy 38~(36) (2013) 16656--16664.

\bibitem{zheng2009cr}
Y.~Zheng, R.~Ran, Z.~Shao, {Cr} doping effect in {B}-{site} of
  {La}$_{0.75}${Sr}$_{0.25}${MnO}$_3$ on its phase stability and performance as
  an {SOFC} anode, Rare Metals 28 (2009) 361--366.

\bibitem{liu2022improvement}
Y.~Liu, J.~Ma, L.~Dai, W.~Meng, L.~Wang, Improvement of the response
  performance of impedimetric {NO}$_2$ sensor by halogen doping of
  {La}$_{0.75}${Sr}$_{0.25}${CrO}$_{3-\delta}$ sensing electrode, {Sensors and
  Actuators B}{:} {Chemical} 358 (2022) 131516.

\bibitem{machado2023chemical}
P.~Machado, R.~Guzm{\'a}n, R.~J. Morera, J.~Alcal{\`a}, A.~Palau, W.~Zhou,
  M.~Coll, Chemical synthesis of {La}$_{0.75}${Sr}$_{0.25}${CrO}$_3$ thin films
  for {p}-{Type} transparent conducting electrodes, Chemistry of Materials
  35~(9) (2023) 3513--3521.

\bibitem{ZHU2023214873}
Y.~Zhu, K.~Sun, S.~Wu, P.~Zhou, Y.~Fu, J.~Xia, H.-F. Li, A comprehensive review
  on the ferroelectric orthochromates: Synthesis, property, and application,
  Coordination Chemistry Reviews 475 (2023) 214873.

\bibitem{chakraborty2006structural}
K.~R. Chakraborty, S.~Yusuf, P.~Krishna, M.~Ramanadham, A.~Tyagi,
  V.~Pomjakushin, Structural study of {La}$_{0.75}${Sr}$_{0.25}${CrO}$_3$ at
  high temperatures, Journal of Physics: Condensed Matter 18~(37) (2006) 8661.

\bibitem{yusuf2008magnetic}
S.~Yusuf, Magnetic correlations in {oxides}{:} {Neutron} diffraction and
  neutron depolarization study, Pramana 71 (2008) 695--704.

\bibitem{khattak1977structural}
C.~Khattak, D.~Cox, Structural studies of the {(}{La}, {Sr}{)}{Cr}{O}$_3$
  system, Materials Research Bulletin 12~(5) (1977) 463--471.

\bibitem{liu2000mixed}
X.~Liu, W.~Su, Z.~Lu, J.~Liu, L.~Pei, W.~Liu, L.~He, Mixed valence state and
  electrical conductivity of {La}$_{1-x}${Sr}$_x${CrO}$_3$, Journal of Alloys
  and Compounds 305~(1-2) (2000) 21--23.

\bibitem{li2008synthesis}
H.~Li, Synthesis of {CMR} manganites and ordering phenomena in complex
  transition metal oxides, Vol.~4, Forschungszentrum J{\"u}lich, 2008.

\bibitem{karim1979localized}
D.~Karim, A.~Aldred, Localized level hopping transport in
  {La}{(}{Sr}{)}{CrO}$_3$, Physical Review B 20~(6) (1979) 2255.

\bibitem{chakraborty2008magnetic}
K.~R. Chakraborty, S.~Yusuf, A.~Tyagi, Magnetic ordering in
  {La}$_{0.75}${Sr}$_{0.25}${CrO}$_3${:} {A} neutron diffraction study, Journal
  of Magnetism and Magnetic Materials 320~(6) (2008) 1163--1166.

\bibitem{devi1992preparation}
P.~S. Devi, M.~S. Rao, Preparation, structure, and properties of
  {strontium}{-}{doped} lanthanum {chromites}{:} {La}$_{1-x}${Sr}$_x${CrO}$_3$,
  Journal of Solid State Chemistry 98~(2) (1992) 237--244.

\bibitem{li2009crystal}
H.-F. Li, Y.~Su, Y.~Xiao, J.~Persson, P.~Meuffels, T.~Brueckel, Crystal and
  magnetic structure of {single}{-}{crystal} {La}$_{1-x}${Sr}$_x${MnO}$_3$
  (${x} {\approx} {1}{/}{8}$), The European Physical Journal B 67 (2009)
  149--157.

\bibitem{li2007neutron}
H.~Li, Y.~Su, J.~Persson, P.~Meuffels, J.~Walter, R.~Skowronek, T.~Br$\ddot{\rm
  u}$ckel, {Neutron}{-}{diffraction} study of structural transition and
  magnetic order in orthorhombic and rhombohedral
  {La}$_{7/8}${Sr}$_{1/8}${Mn}$_{1-\gamma}${O}$_{3 + \delta}$, Journal of
  {Physics}{:} Condensed Matter 19~(17) (2007) 176226.

\bibitem{zhu2019method}
Y.~Zhu, S.~Wu, Z.~Tang, H.-F. Li, A method of centimeter-sized single crystal
  growth of chromate compounds and related storage device, China Patent No. ZL
  2019 1 1281088.8 (2021).

\bibitem{rodriguez1993recent}
J.~Rodr{\'\i}guez-Carvajal, Recent advances in magnetic structure determination
  by neutron powder diffraction, {Physica} {B}{:} {Condensed Matter} 192~(1-2)
  (1993) 55--69.

\bibitem{ouladdiaf2006orientexpress}
B.~Ouladdiaf, J.~Archer, G.~McIntyre, A.~Hewat, D.~Brau, S.~York,
  Orientexpress{:} {A} new system for laue neutron diffraction, {Physica}
  {B}{:} {Condensed Matter} 385 (2006) 1052--1054.

\bibitem{zhu2022crystal}
Y.~Zhu, J.~Xia, S.~Wu, K.~Sun, Y.~Yang, Y.~Zhao, H.~W. Kan, Y.~Zhang, L.~Wang,
  H.~Wang, et~al., Crystal growth engineering and origin of the weak
  ferromagnetism in antiferromagnetic matrix of orthochromates from te orbital
  hybridization, iScience 25~(4) (2022) 104111.

\bibitem{matsunaga2008analysis-2}
Y.~Matsunaga, H.~Kawaji, T.~Atake, H.~Takahashi, T.~Hashimoto, Analysis of
  structural and magnetic phase transition behaviors of
  {La}$_{1-x}${Sr}$_x${Cr}{O}$_3$ by measurement of heat capacity with thermal
  relaxation technique, Thermochimica Acta 474~(1-2) (2008) 57--61.

\bibitem{mazumdar2021structural}
D.~Mazumdar, I.~Das, Structural, magnetic, and magnetocaloric properties of the
  multiferroic host double perovskite compound {Pr}$_2${FeCrO}$_6$, Physical
  Chemistry Chemical Physics 23~(9) (2021) 5596--5606.

\bibitem{li2016possible}
H.~Li, Possible ground states and parallel {magnetic}{-}{field}{-}{driven}
  phase transitions of collinear antiferromagnets, npj Computational Materials
  2~(1) (2016) 1--8.

\bibitem{li2014incommensurate}
H.~Li, C.~Zhang, A.~Senyshyn, A.~Wildes, K.~Schmalzl, W.~Schmidt, M.~Boehm,
  E.~Ressouche, B.~Hou, P.~Meuffels, et~al., Incommensurate antiferromagnetic
  order in the {manifoldly}{-}{frustrated} {SrTb}$_2${O}$_4$ with transition
  temperature up to {4.28} k, Frontiers in Physics 2 (2014) 42.

\bibitem{zhu2020crystalline}
Y.~Zhu, Y.~Fu, B.~Tu, T.~Li, J.~Miao, Q.~Zhao, S.~Wu, J.~Xia, P.~Zhou, A.~Huq,
  et~al., Crystalline and magnetic structures, magnetization, heat capacity,
  and anisotropic magnetostriction effect in a {yttrium}{-}{chromium} oxide,
  Physical Review Materials 4~(9) (2020) 094409.

\bibitem{li2012possible}
H.~Li, Y.~Xiao, B.~Schmitz, J.~Persson, W.~Schmidt, P.~Meuffels, G.~Roth,
  T.~Br{\"u}ckel, Possible {magnetic}{-}{polaron}{-}{switched} positive and
  negative magnetoresistance in the {GdSi} single crystals, Scientific Reports
  2~(1) (2012) 750.

\bibitem{zhu2020high}
Y.~Zhu, S.~Wu, B.~Tu, S.~Jin, A.~Huq, J.~Persson, H.~Gao, D.~Ouyang, Z.~He,
  D.-X. Yao, et~al., {High}{-}{temperature} magnetism and crystallography of a
  {YCrO}$_3$ single crystal, Physical Review B 101~(1) (2020) 014114.

\bibitem{wagner1979handbook}
C.~D. Wagner, Handbook of {X}{-}{ray} photoelectron {spectroscopy}{:} A
  reference book of standard data for use in {X}{-}{ray} photoelectron
  spectroscopy, Vol. Physical Electronics Division, Eden Prairie, {MN}{:}
  {Perkin}{-}{Elmer} {Corporation}, 1979.

\bibitem{Patrycja2018}
P.~Maku{{\l}}a, M.~Pacia, W.~Macyk, How to correctly determine the band gap
  energy of modified semiconductor photocatalysts based on {UV}{–}{Vis}
  spectra, The Journal of Physical Chemistry Letters 9~(23) (2018) 6814--6817.

\bibitem{tezuka1998magnetic}
K.~Tezuka, Y.~Hinatsu, A.~Nakamura, T.~Inami, Y.~Shimojo, Y.~Morii, Magnetic
  and neutron diffraction study on perovskites {La}$_{1-x}${Sr}$_x${Cr}{O}$_3$,
  Journal of Solid State Chemistry 141~(2) (1998) 404--410.

\bibitem{tezuka1998magnetic-2}
K.~Tezuka, Y.~Hinatsu, K.~Oikawa, Y.~Shimojo, Y.~Morii, Studies on magnetic
  properties of {La}$_{0.95}${Sr}$_{0.05}${Cr}{O}$_3$ and
  {La}$_{0.85}${Sr}$_{0.15}${Cr}{O}$_3$ by means of powder neutron diffraction,
  Journal of Physics{:} {Condensed} Matter 12~(18) (2000) 4151--4160.

\bibitem{nakamura2005analysis}
F.~Nakamura, Y.~Matsunaga, N.~Ohba, K.~Arai, H.~Matsubara, H.~Takahashi,
  T.~Hashimoto, Analysis of magnetic and structural phase transition behaviors
  of {La}$_{1-x}${Sr}$_x${Cr}{O}$_3$ for preparation of phase diagram,
  Thermochimica Acta 435~(2) (2005) 222--229.

\bibitem{matsunaga2008analysis}
Y.~Matsunaga, F.~Nakamura, H.~Takahashi, T.~Hashimoto, Analysis of relationship
  between magnetic property and crystal structure of
  {La}$_{1-x}${Sr}$_x${Cr}{O}$_3$ ({x} {=} 0.13, 0.15), Solid State
  Communications 145~(9-10) (2008) 502--506.

\bibitem{SILVAJR2022139278}


\bibitem{biswas2014magnetization}
S.~Biswas, M.~H. Khan, S.~Pal, Magnetization reversal and ferrimagnetism in
  {Pr}$_{1-x}${Nd}$_x${MnO}$_3$, Bulletin of Materials Science 37 (2014)
  809--813.

\bibitem{Khomchenko2008}
V.~A. Khomchenko, I.~O. Troyanchuk, R.~Szymczak, H.~Szymczak, Negative
  magnetization in {La}$_{0.75}${Nd}$_{0.25}${Cr}{O}$_3$ perovskite, Journal of
  Materials Science 43~(16) (2008) 5662--5665.

\bibitem{biswas2018negative}
S.~Biswas, S.~Pal, Negative magnetization in perovskite {RTO}$_3$ ({R} =
  rare-earth, {T} = {Cr}{/}{Mn}), Reviews on Advanced Materials Science 53~(2)
  (2018) 206--217.

\bibitem{kumar2015phenomenon}
A.~Kumar, S.~Yusuf, The phenomenon of negative magnetization and its
  implications, Physics Reports 556 (2015) 1--34.

\bibitem{yoshii2001magnetic}
K.~Yoshii, Magnetic properties of perovskite {GdCrO}$_3$, Journal of Solid
  State Chemistry 159~(1) (2001) 204--208.

\bibitem{jaiswal2010magnetic}
A.~Jaiswal, R.~Das, K.~Vivekanand, T.~Maity, P.~M. Abraham, S.~Adyanthaya,
  P.~Poddar, Magnetic and dielectric properties and raman spectroscopy of
  {GdCrO}$_3$ nanoparticles, Journal of Applied Physics 107~(1) (2010).

\bibitem{cao2014temperature}
S.~Cao, H.~Zhao, B.~Kang, J.~Zhang, W.~Ren, Temperature induced spin switching
  in {SmFeO}$_3$ single crystal, Scientific reports 4~(1) (2014) 5960.

\end{thebibliography}

\clearpage

\clearpage

\clearpage

\begin{table}[t]
\small
\caption{Refined room-temperature structural details of the La$_{0.75}$Sr$_{0.25}$Cr$_{0.90}$O$_{3-\delta}$ single crystal. These structural parameters include lattice constants, unit-cell volume ($V$), atomic positions, isotropic thermal parameters ($B$), and goodness of fit. The Wyckoff sites of all atoms were listed. We constrained equal $B$ values for La and Sr ions during refinements. The numbers in parentheses are the estimated standard deviations of the last significant digit.}
\label{latticep}
%\begin{ruledtabular}
\setlength{\tabcolsep}{12mm}{}
\renewcommand{\arraystretch}{1.0}
\begin{tabular}{ll}
\hline
\hline
\multicolumn{2}{c}{Pulverized La$_{0.75}$Sr$_{0.25}$Cr$_{0.90}$O$_{3-\delta}$ Single Crystal} \\
\hline
Crystal System                                 &Trigonal                                      \\
Space group                                    &$R\bar{3}c$ (No. 167)                         \\
$Z$                                            &6                                             \\
\hline
$a (= b)$ (\AA)                                &5.49754(6)                                    \\
$c$ (\AA)                                      &13.29480(2)                                   \\
$\alpha (= \beta)$ ($^{\circ}$)                &90                                            \\
$\gamma$ ($^{\circ}$)                          &120                                           \\
Unit-cell $V$ (\AA$^3$)                        &347.977(7)                                    \\
\hline
La/Sr                                          &6$a$                                          \\
$x$                                            &0                                             \\
$y$                                            &0                                             \\
$z$                                            &0.25                                          \\
$B$ (\AA$^2$)                                  &1.99(2)                                       \\
\hline
Cr                                             &6$b$                                          \\
$x$                                            &0                                             \\
$y$                                            &0                                             \\
$z$                                            &0                                             \\
$B$ (\AA$^2$)                                  &2.25(3)                                       \\
\hline
O                                              &18$e$                                         \\
$x$                                            &0.4461(4)                                     \\
$y$                                            &0                                             \\
$z$                                            &0.25                                          \\
$B$ (\AA$^2$)                                  &1.36(6)                                       \\
\hline
$R_\textrm{p}$                                 &6.29                                          \\
$R_\textrm{wp}$                                &8.11                                          \\
$R_\textrm{exp}$                               &6.42                                          \\
$\chi^{2}$                                     &1.60                                          \\
\hline
\hline
\end{tabular}
%\end{ruledtabular}
\end{table}

\clearpage

\clearpage

\clearpage

\clearpage

\clearpage

\clearpage

\clearpage

\clearpage

\begin{table*}[t]
\small
\caption{Summary of the crystal and potential magnetic structures, as well as phase-transition properties, for both the single-crystal La$_{0.75}$Sr$_{0.25}$Cr$_{0.90}$O$_{3-\delta}$ (investigated in this study) and the polycrystalline La$_{1-x}$Sr$_x$CrO$_3$ (x = 0.15, 0.25; sourced from literature) compounds. The notations used in the summary are as follows: S. = single crystal, P. = polycrystal, Atmos. = synthesizing atmosphere, Crys. = crystal system, $T =$ temperature, SR = spin reorientation, FM = ferromagnetic component formed by AFM canting, Ref. = reference, and TS = this study.}
\label{Comp}
\setlength{\tabcolsep}{1.5mm}{}
\renewcommand{\arraystretch}{1.0}
%\begin{ruledtabular}
\begin{tabular}{lllllll}
\hline
\hline
Compound                         &Atmos.                                        &Crys.                                       &Transition $T$     &Transition type                                     &Spin structure                    &Ref.                                                      \\
                                 &                                              &                                            &(K)                &                                                    &                                  &                                                          \\
\hline
S. La$_{0.75}$Sr$_{0.25}$Cr$_{0.90}$O$_{3-\delta}$ &O$_2$                                         &$R\overline{3}c$                            &271.80(1)          &PM to canted-AFM                                    &                                  &TS                                                        \\
                                 &                                              &                                            &155.01(1)          &Structure/SR                                        &                                  &TS                                                        \\
                                 &                                              &                                            &117.94(1)          &AFM1 formed                                         &                                  &TS                                                        \\
                                 &                                              &                                            &34.98(1)           &FM/SR                                               &                                  &TS                                                        \\
                                 &                                              &                                            &21.50(1)           &AFM2 formed                                         &                                  &TS                                                        \\
\hline
P. La$_{0.75}$Sr$_{0.25}$CrO$_3$ &Air                                           &$Pbnm$, $R\overline{3}c$                    &250--300           &PM to AFM                                           &$G_y$ AFM                         &\cite{chakraborty2008magnetic}                            \\
P. La$_{0.75}$Sr$_{0.25}$CrO$_3$ &Air                                           &$R\overline{3}c$                            &242--247           &PM to canted-AFM                                    &                                  &\cite{matsunaga2008analysis-2}                            \\
P. La$_{0.75}$Sr$_{0.25}$CrO$_3$ &O$_2$                                         &$R\overline{3}c$                            &$\sim$ 250         &PM to AFM                                           &                                  &\cite{tezuka1998magnetic}                                 \\
\hline
P. La$_{0.85}$Sr$_{0.15}$CrO$_3$ &O$_2$ \cite{tezuka1998magnetic}               &$R\overline{3}c$                            &266--271           &PM to AFM                                           &$G_{[111]}$ AFM                   &\cite{tezuka1998magnetic,tezuka1998magnetic-2}            \\
 							     &Air \cite{tezuka1998magnetic-2}               &$Pnma$                                      &190                &Structure $R\overline{3}c$ to $Pnma$                &                                  &\cite{tezuka1998magnetic,tezuka1998magnetic-2}            \\
 							     &                                              &                                            &                   &Magnetic                                            &$G_z$ (AFM), $F_y$ (FM)           &\cite{tezuka1998magnetic,tezuka1998magnetic-2}            \\
                                 &                                              &$Pnma$                                      &160                &Spin reorientation                                  &$G_z$ AFM                         &\cite{tezuka1998magnetic,tezuka1998magnetic-2}            \\
                                 \hline
\hline
\end{tabular}
%\end{ruledtabular}
\end{table*}

\clearpage

\end{document}